# Modulating lipid membrane morphology by dynamic DNA origami networks

*Juanjuan Yang‡,[a] [b] Kevin Jahnke† ‡,[c] Ling Xin‡,[b] Xinxin Jing, [b] Pengfei Zhan,[b] Andreas Peil,[b] Alessandra Griffo,[c] Marko Škugor,[b] Donglei Yang,[a] Sisi Fan,[b] Kerstin Göpfrich,*[c][d] Hao Yan,*[e] Pengfei Wang,*[a] and Na Liu*[b][f]*

[a] Institute of Molecular Medicine, Department of Laboratory Medicine, Shanghai Key Laboratory for Nucleic Acid Chemistry and Nanomedicine, Renji Hospital, School of Medicine, Shanghai Jiao Tong University Shanghai 200127, P. R. China.
[b] 2. Physics Institute, University of Stuttgart, Pfaffenwaldring 57, 70569, Stuttgart, Germany.
[c] Biophysical Engineering Group, Max Planck Institute for Medical Research Heidelberg, Jahnstr. 29, 69120 Heidelberg, Germany.
[d] Center for Molecular Biology (ZMBH), Heidelberg University, Im Neuenheimer Feld 329, 69120 Heidelberg, Germany.
[e] School of Molecular Sciences and Center for Molecular Design and Biomimetics at Biodesign Institute, Arizona State University, Tempe, AZ, USA.
[f] Max Planck Institute for Solid State Research, Heisenbergstrasse 1, D-70569 Stuttgart, Germany.

ABSTRACT. Membrane morphology and its dynamic adaptation regulate many cellular functions, which are often mediated by membrane proteins. Advances in DNA nanotechnology have enabled the realization of various protein-inspired structures and functions with precise control at the nanometer level, suggesting a viable tool to artificially engineer the membrane morphology. In this work, we demonstrate a DNA origami cross (DOC) structure that can be anchored onto giant unilamellar vesicles (GUVs) and subsequently polymerized into micron-scale reconfigurable one-dimensional (1D) chains or two-dimensional (2D) lattices. Such DNA origami-based networks can be switched between left-handed (LH) and right-handed (RH) conformations by DNA fuels and exhibit potent efficacy in remodeling the membrane curvatures of GUVs. This work sheds light on designing hierarchically-assembled dynamic DNA systems for the programmable modulation of synthetic cells for useful applications.

Biological membranes are dynamic and fluidic entities, whose morphologies play an essential role in many cellular functions, such as cell division, cell migration, signal transduction, and functional intracellular compartmentalization[1]. The mechanisms for membrane remodeling associated with membrane-bound proteins generally include (i) scaffolding by proteins with curved architectures, (ii) insertion of amphipathic helices, and (iii) protein crowding[2, 3]. Recent advances in DNA nanotechnology have enabled the creation of complex DNA structures with a variety of protein-inspired functions and structural control at the nanometer level[4-7]. This renders DNA nanotechnology a unique tool to engineer membrane morphologies from the bottom up[8-10]. There have been continuous experimental efforts towards membrane sculpting by DNA structures[11-15]. For instance, curved DNA nanostructures that emulated BAR domain proteins were realized to induce membrane deformation[16]. Czogalla and colleagues constructed DNA origami blocks, which led to vesicle flattening by lateral polymerization[17]. Closely related to this work, Franquelim and colleagues implemented DNA origami rods, which could be polymerized into filaments to promote membrane protrusions[18].

As a matter of fact, many membrane-bound proteins that are involved in cell morphology reshaping are often found in chiral configurations[19]. For example, dynamins that assemble on the neck of a clathrin-coated membrane bud form a helix. Also, BAR domain proteins are chiral and bend the membrane via scaffolding. Although the role of such chirality remains elusive, it is still instructive to construct synthetic systems with inherent chirality and examine their capability of membrane remodeling. In particular, DNA-based nanostructures can be rationally designed with switchable chirality and programmable functions under simple experimental settings, while control over these features is generally difficult to achieve in complex protein systems. In this work, we demonstrate a membrane-bound DNA origami cross (DOC) chiral structure, which is capable of both polymerization and reconfiguration to engineer the membrane morphology of giant unilamellar vesicles (GUVs). More specifically, the DOCs can be polymerized into reconfigurable 1D filament-like chains or 2D lattices by DNA hybridization[20, 21]. The reconfiguration of the microscale DNA architectures is driven by strand displacement reactions[22-24]. After anchoring the DOCs onto GUVs (Figure 1a), we show that their polymerization can result in membrane deformation to different extents depending on the polymerization degree of the DOC networks.

The DOC was designed using the CaDNAno[25] software (Figure S1). As shown in Figure 1b, the DOC consists of two linked helix bundles, denoted as Up (B) and Down (A), respectively. The 12-helix Up bundle has a length of 74.5 nm and a width of 15 nm. It contains two rafts at its two ends, resulting in a total height of 12.5 nm. The 16-helix Down bundle also has a length of 74.5 nm and a width of 15 nm. The inclusion of the rafts on the Up bundle ensures the bottoms of the two bundles lie flat on the same surface. Single-stranded DNA captures are extended from the bottoms of the two bundles for binding with the cholesterol-modified complementary DNA anchors on lipid membranes[26-28]. The two ends of the DOC are modified, so that upon the addition of DNA connector strands (connector A or/and connector B), the DOC monomers can be polymerized into 1D chains or 2D lattices by DNA hybridization (Figure 1c). To form 1D chains, polymerization can take place along the Up bundles using connector B (Figure S2a) or the Down bundles using connector A (Figure S2b). The angle (45° or 135°) between the two bundles in the DOC is controlled by two sets of DNA locks (lock LH and lock RH), which can respond to toehold-mediated strand displacement reactions by the addition of corresponding switching strands (Figure 1c). The DOC can thus be switched to a left-handed monomer (45°, LH-M) or a right-handed monomer (135°, RH-M) as shown in Figure 1b. The chirality of the 1D chains or the 2D lattices formed by the DOC polymerization are modulated *via* the same strategy (Figure 1c).

We first experimentally demonstrate the formation of the LH and RH DOC monomers. Purified DOCs by agarose gel electrophoresis (Figure S3) were characterized using atomic force microscopy (AFM) and transmission electron microscopy (TEM). As shown in Figures 2a and 2b (see also Figures S4-S5), the AFM and TEM images confirm the successful assembly of the RH-Ms and LH-Ms, which are locked to 135° and 45°, respectively. To enable the polymerization, the respective connector strands were added to the DOC monomers for the assembly of 1D chains (Figures 2c-2f and Figures S6-S9) and 2D lattices (Figures 2g and 2h) through DNA hybridization[29]. The mean length of the 1D-Up chains in both handedness, i.e., RH-1D-Up (Figure 2c) and LH-1D-Up (Figure 2d), is 354.1 ± 40.3 nm (Figure S9), whereas the 1D-Down chains that are polymerized along the Down bundles in both handedness, i.e., RH-1D-Down (Figure 2e) and LH-1D-Down (Figure 2f), are 724.5 ± 80.2 nm (Figure S8). The polymerization length along the Down bundles is thus significantly longer than that along the Up bundles (see also Figure S9). This is likely due to the intrinsic twist and floppiness within the Up bundle (Figure S10), which impedes the end-to-end connection between them for the long polymerization[30, 31]. The 2D lattices formed by linking along both the Up and Down bundles are shown in Figures 2g-2h, and Figure S11. Similar to the 1D chains, the 2D lattices in both handedness exhibit a higher preference to growing along the Down bundle direction.

We next sought to demonstrate chirality switching of the 1D chains and 2D lattices. The RH-1D-Down chains were dispersed on mica and imaged using AFM as shown in Figure 3a. The height profile measured along the Up bundle (see the white line) reveals that it is the bottom view of the chain, in which the facing-up rafts introduce abrupt height changes between the Up and Down bundles. The switching experiment from the RH-1D-Down to LH-1D-Down chains was carried out in solution. Each DNA lock consists of three parts (Figure S12): (i) a double-stranded region to control the distance, (ii) a complementary region to lock the two bundles and (iii) a toehold region to conduct toehold-mediated strand displacement reactions. Upon the addition of the switching strands, the locks RH were open and the DOCs were subsequently switched to LH by activation of the locks LH. The LH-1D-Down chains were then dispersed on mica and imaged using AFM. The height measurement along the Up bundle exhibits rather a flat profile, indicating that it is the top view of the chain. Dynamic switching in cycles can be found in Figure S13. Similarly, the chirality switching can also be successfully executed for the 2D lattices from RH-2D to LH-2D as shown in Figure 3b.

To anchor the DOCs onto the outer membranes of GUVs (DOPC with 0.5% Atto488-labled DOPE), five capture strands are extended from the DOC's bottom, one on each raft from the Up bundle and three from the Down bundle (Figures S14 and S15). These capture strands are complementary to the cholesterol-modified DNA handles that self-assemble into the membrane of preformed GUVs through hydrophobic interactions (Figure 4a). For visualization, the DOCs and DOPE lipids were modified with Atto647 and Atto488 dyes, respectively. Purified LH-Ms (3 nM) were incubated with GUVs, which were functionalized with cholesterol-modified DNA anchors. Confocal laser-scanning microscopy confirms the binding of the DOC monomers onto the GUVs, as the fluorescence signals of the lipids (green) and DOCs (red) colocalize as shown in Figure 4b. We next investigate the diffusability of the DOCs on the GUVs using fluorescence recovery after photobleaching (FRAP). LH-Ms with 10 mM $Mg^{2+}$ was applied. After photobleaching the localized region of the GUV as shown in Figure 4c, the lipid fluorescence signals are recovered very fast, whereas the DOC fluorescence signals only show a slight increase (Figure 4c). This indicates that the lipid membrane is diffusive (diffusion coefficient = 2.39± 0.28 $\mu m^2/s$), while the DOCs exhibit a low mobility on the GUV (diffusion coefficient =1.4 ± 0.65 $\mu m^2/s$). It is noteworthy that upon the addition of 75 mM NaCl, the mobility of the DOCs can be greatly enhanced[32, 33] (Figure S16).

To induce the polymerization of the membrane-bound LH-Ms, DNA connector strands were added into the solution and incubated with the GUVs. The FRAP results of the LH-Ms, 1D chains (LH-1D-Up and LH-1D-Down) and 2D lattices (LH-2D) are shown in Figure 4d. The retrieved diffusion coefficients are displayed in Figure 4e. The diffusion coefficients of the lipids in the cases of the 1D chains and 2D lattices are approximately in the same range as that of the bare GUVs. The diffusion coefficients of the LH-1D-Down and LH-2D are much lower than those of the LH-1D-Up and LH-Ms. The polar lipid head groups interact with the highly charged membrane-absorbed DNA origami. Through this interaction, the non-diffusive DNA origami can reduce the lipid diffusion. The larger the DNA assemblies, the stronger this effect is expected to be. This corroborates our previous observations that the 1D-Up chains have a low polymerization degree as presented in Figure 2.

Finally, we evaluate the influence of the DOC polymerization on the membrane remodeling of the GUVs. In order to modulate the GUV's surface tension and to provide sufficient surface area for deformation, the vesicles were osmotically deflated by adding 300 mM sucrose into the solution. In the time lapse study, the GUVs functionalized with different DOCs (1 nM) were monitored for 40 mins after the GUV deflation. We chose a DOC concentration of 1 nM in the experiments, because it allowed for efficient DOC binding with good fluorescence signals, meanwhile it did not induce GUV deformations before the DOC polymerization (Figure S17). For the GUVs covered by LH-Ms, the GUV shape remains almost spherical (Figure 5a) and thus the corresponding circularity approaches 1.0 (Figures 5e and S18). In contrast, the GUVs incubated with DNA connectors exhibit evident deformations to different extents (Figures 5b-5d). More specifically, the LH-1D-Up chains deform the GUVs the most moderately, whereas the LH-2D lattices introduce the highest deformation of the GUVs. The degree of the GUV deformation is in close relationship with the degree of the DOC polymerization. That is, the longer the chains or the larger the lattices grow, the stronger the GUVs are deformed (Figure S19-S21). The corresponding circularity data in Figures 5f-5h quantitatively validate the observations from the confocal fluorescence results in Figures 5b-5d. We also examined the reconfiguration of the 2D-DOCs on GUVs after deflation and triggered the configuration change of the 2D-DOCs by the addition of Cy3-labeled switching strands (Figure S22).

In conclusion, we have demonstrated a reconfigurable DNA origami building unit that can be bound to lipid membranes and polymerized upon the addition of the connector strands into 1D chains and 2D lattices to remodel GUVs. The deformation degree of the GUVs correlates significantly with the polymerization degree of the DNA origami building unit on the GUVs. Cell morphology is largely controlled by membrane-bound proteins, which can drive membrane curvature, tabulation, and many peculiar shapes[34]. Mimicking membrane deformations using dynamic DNA nanostructures with tunability and programmability could help us to gain insights into the fundamental cellular processes, such as endocytosis, exocytosis, and develop engineered tools to modulate them in the future. Beyond that, it would be interesting to develop DNA nanomachines that can generate constriction forces to mediate the fission of GUVs or even morphology manipulation of living cells.

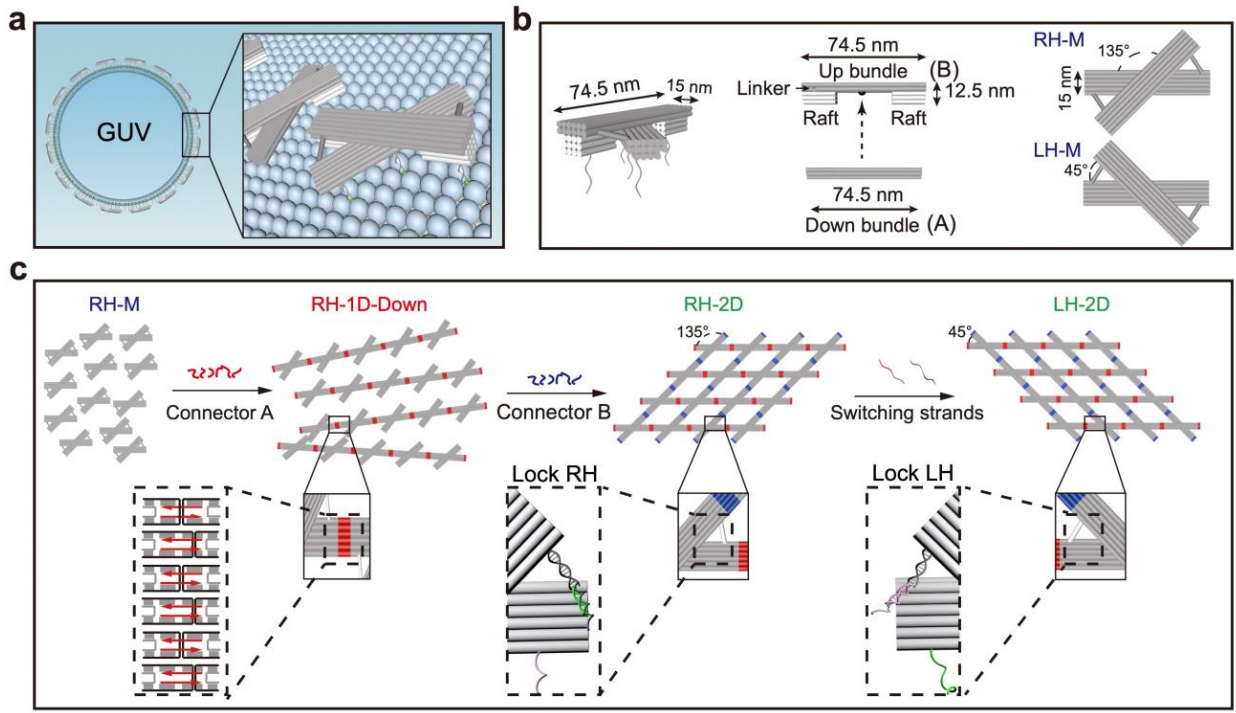

**Figure 1.** Reconfigurable DNA origami networks for lipid membrane modulation. (**a**) Membrane-bound DNA origami cross (DOC) structures on a giant unilameller lipid vesicle (GUV). (**b**) Schematic of the DOC monomer. The DOC consists of linked Up and Down bundles. The Up bundle is designed with two raft extensions at the two ends. The Down bundle is linked to the Up bundle by two single stranded scaffold connections at their pivot point. The dimension of the Up bundle is 74.5 nm in length, 15 nm in width and 12.5 nm in height. The angle between the two bundles can be 135° to form a right-handed monomer (RH-M) or 45° to form a left-handed monomer (LH-M). DNA capture strands are extended from the bottom of the DOC, and they are complementary to the cholesterol handles anchored on the GUVs. (**c**) Hierarchical assembly and reconfiguration of the DOC networks. The DOC monomers can be connected along the Down bundles using connector strands A to form a 1D-Down chain, which can then be further connected along the Up bundles to form a 2D lattice (RH-2D) by connector strands B. RH-2D can be switched to LH-2D and vice versa controlled by lock RH and lock LH via toehold-mediated strand displacement reactions.

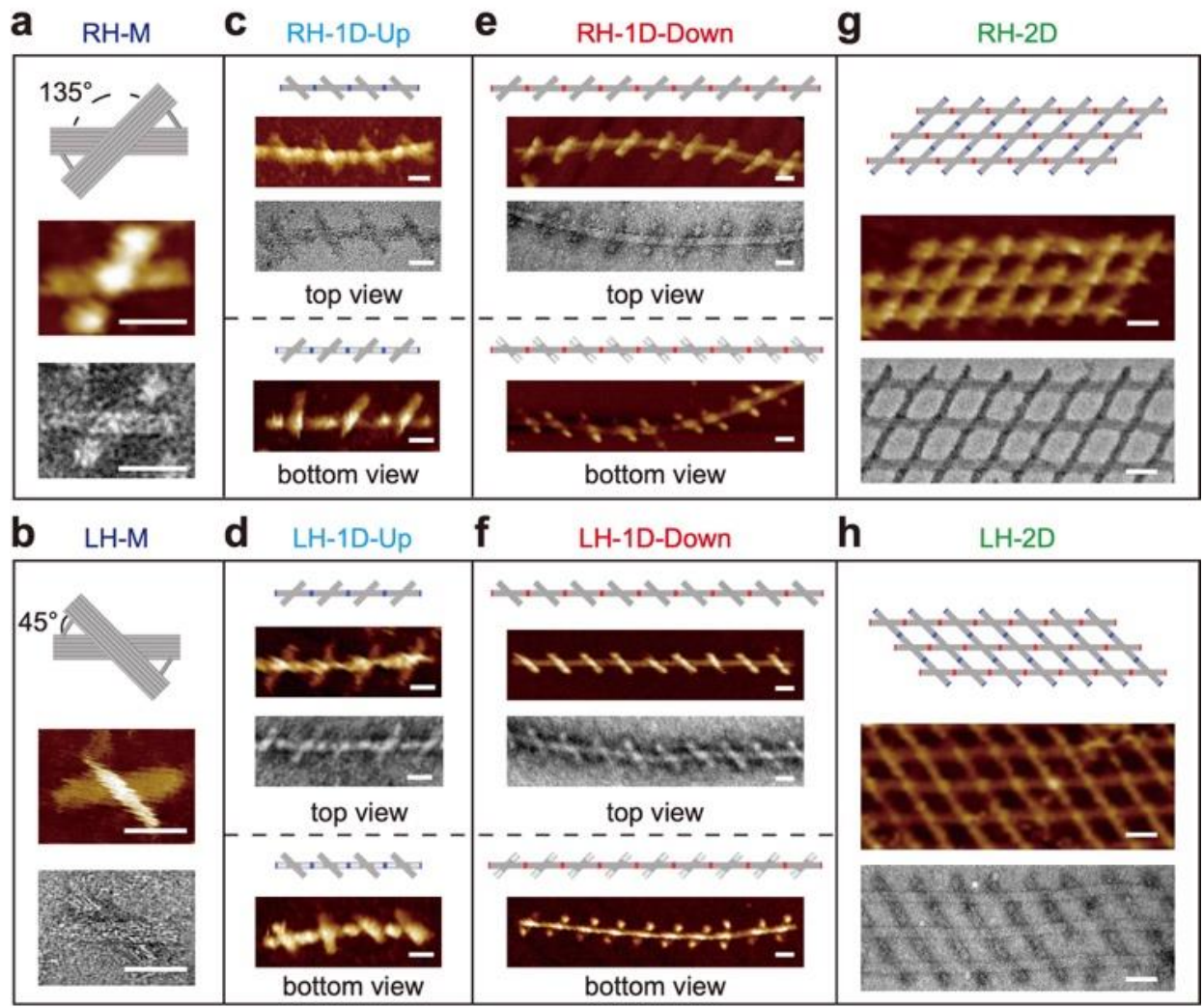

**Figure 2.** AFM and TEM characterizations of the DOC monomers and higher-order 1D chains and 2D lattices with controlled chirality. (a, b) DOC monomers (RH-M, LH-M). (c, d) 1D chains polymerized *via* the Up bundles (RH-1D-Up, LH-1D-Up). (e, f) 1D chains polymerized *via* the Down bundles (RH-1D-Down, LH-1D-Down). (g, h) 2D lattices polymerized *via* both the Up and Down bundles (RH-2D, LH-2D). Scale bars: 50 nm.

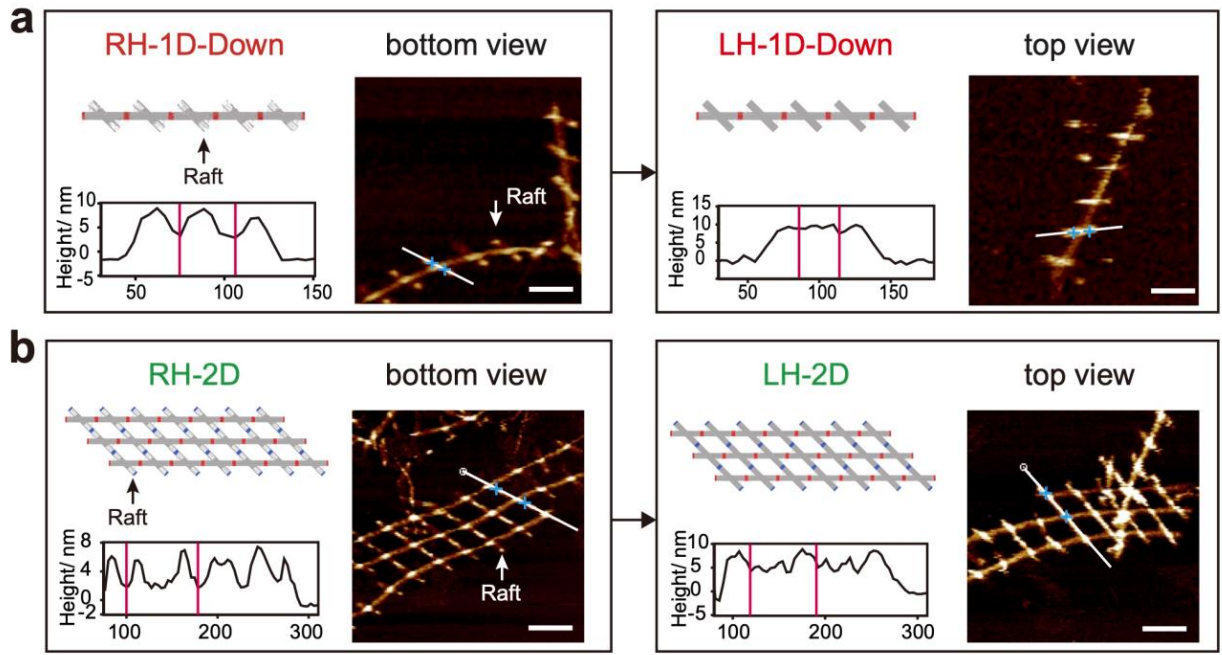

**Figure 3.** Reconfiguration of the 1D chains and 2D lattices characterized by AFM. (a) Configuration change of the 1D chains from RH to LH. (b) Configuration change of the 2D lattices from RH to LH. AFM images, Scale bars: 100 nm. Insets: height measurements along the highlighted white lines.

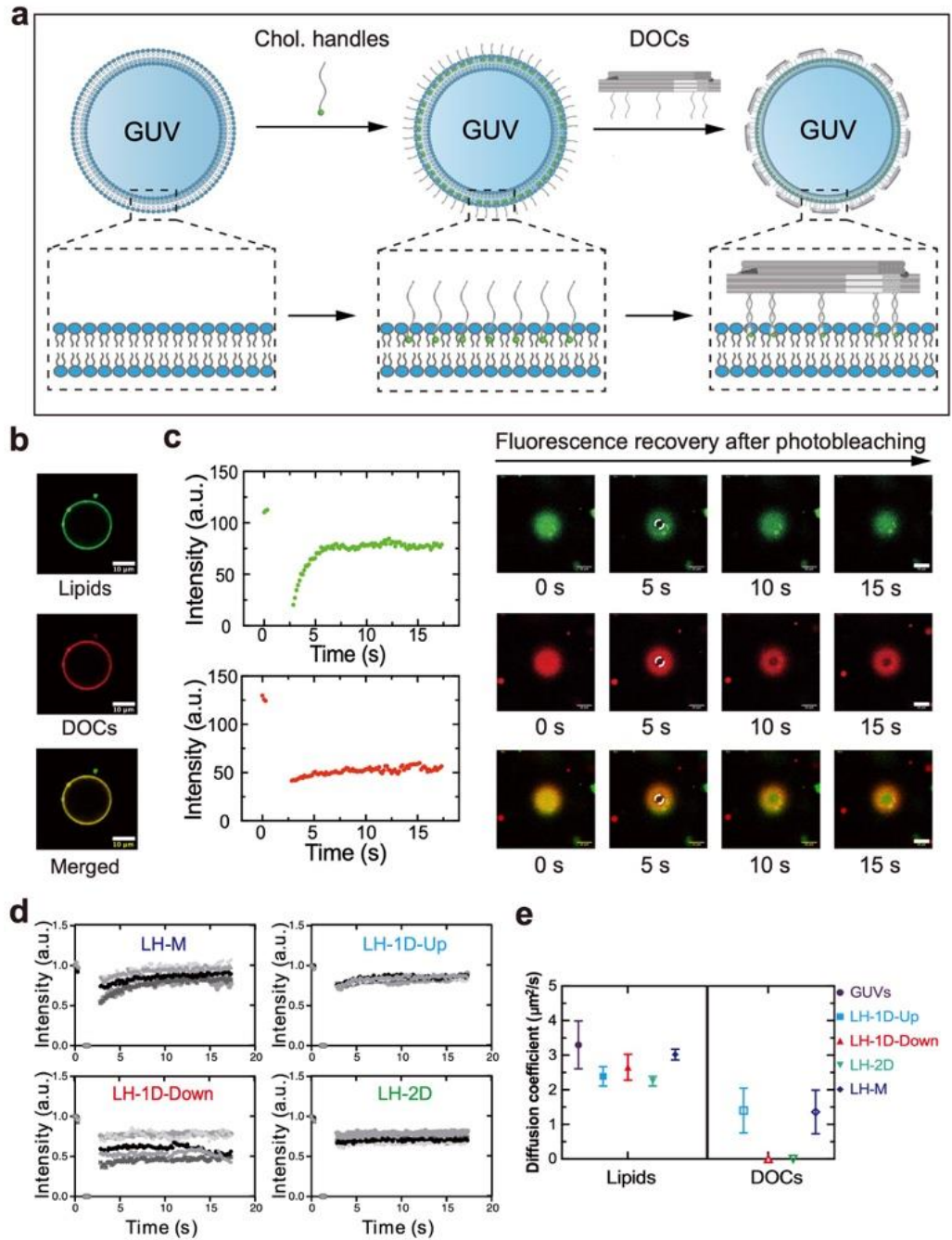

**Figure 4.** Mobility of the membrane-bound DOCs on GUVs. (a) Schematic of the anchoring mechanism of the DOC structures onto the GUV membranes. (b) Confocal microscopy images of a DOPC GUV labeled with 0.5% Atto488-DOPE and the bound DOC monomers labeled with Atto647 in the equatorial plane. (c) FRAP experiments reveal a slow diffusion of the DOC monomers on the GUVs in the presence of divalent cations. (d) FRAP experiments with different DOC structures on GUVs. (e) Summary chart of the diffusion coefficients for lipids and different DOC structures retrieved from the FRAP experiments (N≥5).

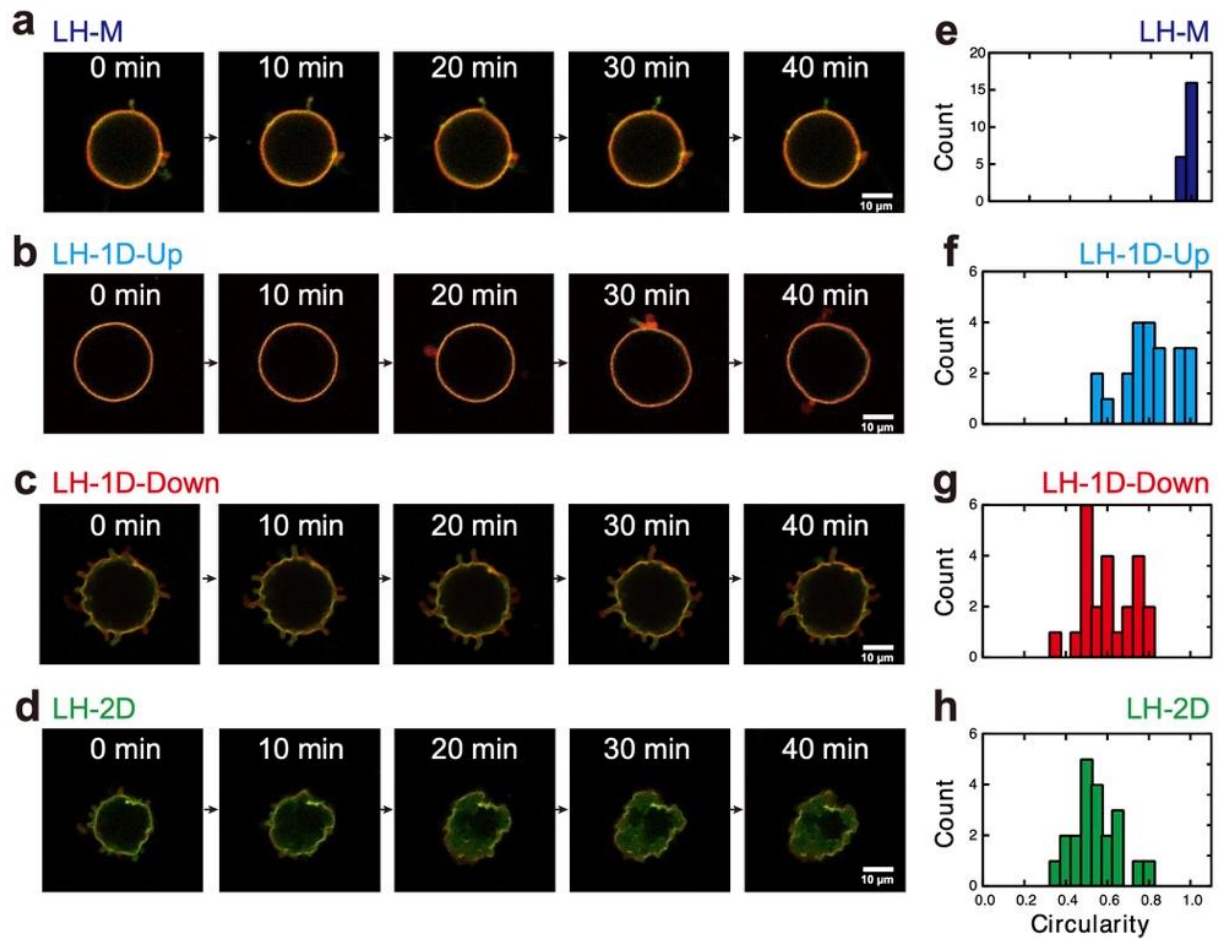

**Figure 5.** Membrane deformation by polymerization of the DOCs on the GUVs after osmotic deflation. Osmolarity ratio: c/c0=1.5. Time lapse tracking of a single GUV bound by (a) LH-Ms, (b) LH-1D-Up chains, (c) LH-1D-Down chains, (d) LH-2D lattices. The circularity distribution of the GUVs after 24 hours incubation is plotted for (e) LH-Ms, (f) LH-1D-Up chains, (g) LH-1D-Down chains, and (h) LH-2D lattices (N=20).

## ASSOCIATED CONTENT

**Supporting Information**.

The following files are available free of charge.

Experimental, analytical details; TEM, AFM and gel images; Origami sequence information; including Figures S1-S22(PDF)

## AUTHOR INFORMATION

**Corresponding Author**

* Kerstin Göpfrich: kerstin.goepfrich@mr.mpg.de

* Hao Yan: hao.yan@asu.edu

* Pengfei Wang: pengfei.wang@sjtu.edu.cn

* Na Liu: na.liu@pi2.uni-stuttgart.de

**Present Addresses**

†Dr. K. Jahnke Present address: School of Engineering and Applied Sciences Harvard University 9 Oxford St., Cambridge, Massachusetts, USA

**Author Contributions**

L.X., N.L., and P.W. initiated the project. L.X. designed the DNA origami. J.Y. designed and performed the experiments. K.J. prepared the vesicles and performed the fluorescence experiments. J.Y. and K.J. analyzed data and prepared the manuscript. N.L. and P.W. supervised the study and revised the manuscript. All authors reviewed and approved the manuscript.

‡These authors contributed equally.

**Funding Sources**

P.W. and J.Y. acknowledge support from National Natural Science Foundation of China (52061135109). The authors from University of Stuttgart received funding from the German Research Foundation (DFG, 448727036), the European Union's Horizon 2020 research and innovation program under grant agreement No. 964995, and the Baden-Württemberg Stiftung (Internationale Spitzenforschung, BWST-ISF2020-19). N.L. also acknowledges support from the Max Planck Society (Max Planck Fellow). K.J. thanks the Joachim Herz Foundation for financial support. K.G. thanks the Hector Fellow Academy, the Carl Zeiss Foundation and the Deutsche Forschungsgemeinschaft (DFG, German Research Foundation) under Germany's Excellence Strategy via the Excellence Cluster 3D Matter Made to Order (EXC-2082/1 - 390761711).

**Notes**

The authors declare no competing financial interest.

ACKNOWLEDGMENT

We thank Prof. Shuoxing Jiang for advice with the origami design. We thank Shuo Wang, Prof. Stephan Nussberger and Prof. Robin Ghosh for assistance with the GUV production. TEM images were acquired at the Stuttgart Center for Electron Microscopy under the supervision of Marion Kelsch. We thank Burghard Zaklina for the assistance with AFM. We thank Stephan Eisler for the assistance with confocal microscopy performed at the microscopy center of the Stuttgart Research Center for Systems Biology.

# Table of Contents

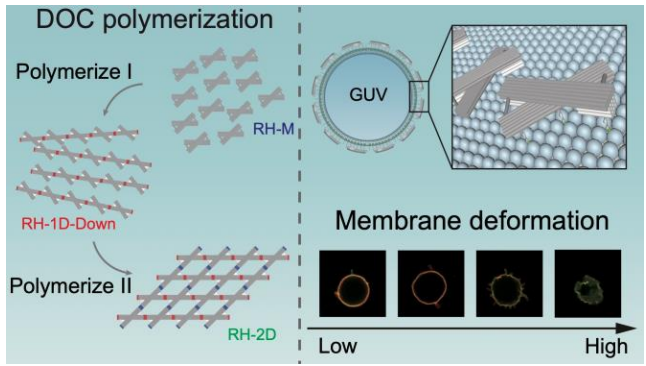

We demonstrate a membrane-bound DNA origami cross structure, which is capable of both polymerization and reconfiguration to engineer the membrane morphology of giant unilamellar vesicles.